%% file: main.tex
\documentclass[
    reprint,
    aps,
    prx,
	amsmath,
	amssymb,
	superscriptaddress,
	longbibliography,
	english
]{revtex4-2}
\usepackage{style} 

\begin{document}

\title{Superconducting Qubit Gates Robust to Parameter Fluctuations}

\author{E.~M.~Wright}
\email{emily.wright@wmi.badw.de}
\affiliation{Technical University of Munich, TUM School of Natural Sciences, Department of Physics, Garching 85748, Germany}
\affiliation{Walther-Mei{\ss}ner-Institut, Bayerische Akademie der Wissenschaften, Garching 85748, Germany}
\affiliation{International Max Planck Research School for Quantum Science and Technology (IMPRS-QST), Max Planck Institute of Quantum Optics, Garching 85748, Germany}

\author{L.~Van Damme}
\affiliation{Technical University of Munich, TUM School of Natural Sciences, Department of Physics, Garching 85748, Germany}
\affiliation{Munich Center for Quantum Science and Technology (MCQST), Schellingstra{\ss}e 4, M\"{u}nchen 80799 , Germany}

\author{N.~J.~Glaser}
\affiliation{Technical University of Munich, TUM School of Natural Sciences, Department of Physics, Garching 85748, Germany}
\affiliation{Walther-Mei{\ss}ner-Institut, Bayerische Akademie der Wissenschaften, Garching 85748, Germany}

\author{A.~Devra}
\affiliation{Technical University of Munich, TUM School of Natural Sciences, Department of Physics, Garching 85748, Germany}
\affiliation{Munich Center for Quantum Science and Technology (MCQST), Schellingstra{\ss}e 4, M\"{u}nchen 80799 , Germany}

\author{F.~A.~Roy}
\affiliation{Walther-Mei{\ss}ner-Institut, Bayerische Akademie der Wissenschaften, Garching 85748, Germany}
\affiliation{Theoretical Physics, Saarland University, Saarbr\"{u}cken 66123, Germany}

\author{J.~Englhardt}
\affiliation{Technical University of Munich, TUM School of Natural Sciences, Department of Physics, Garching 85748, Germany}
\affiliation{Walther-Mei{\ss}ner-Institut, Bayerische Akademie der Wissenschaften, Garching 85748, Germany}

\author{N.~Bruckmoser}
\affiliation{Technical University of Munich, TUM School of Natural Sciences, Department of Physics, Garching 85748, Germany}
\affiliation{Walther-Mei{\ss}ner-Institut, Bayerische Akademie der Wissenschaften, Garching 85748, Germany}

\author{L.~Koch}
\affiliation{Technical University of Munich, TUM School of Natural Sciences, Department of Physics, Garching 85748, Germany}
\affiliation{Walther-Mei{\ss}ner-Institut, Bayerische Akademie der Wissenschaften, Garching 85748, Germany}

\author{A.~Marx}
\affiliation{Technical University of Munich, TUM School of Natural Sciences, Department of Physics, Garching 85748, Germany}
\affiliation{Walther-Mei{\ss}ner-Institut, Bayerische Akademie der Wissenschaften, Garching 85748, Germany}

\author{J.~Schirk}
\affiliation{Technical University of Munich, TUM School of Natural Sciences, Department of Physics, Garching 85748, Germany}
\affiliation{Walther-Mei{\ss}ner-Institut, Bayerische Akademie der Wissenschaften, Garching 85748, Germany}

\author{C.~M.~F.~Schneider}
\affiliation{Technical University of Munich, TUM School of Natural Sciences, Department of Physics, Garching 85748, Germany}
\affiliation{Walther-Mei{\ss}ner-Institut, Bayerische Akademie der Wissenschaften, Garching 85748, Germany}

\author{L.~S\"{o}dergren}
\affiliation{Technical University of Munich, TUM School of Natural Sciences, Department of Physics, Garching 85748, Germany}
\affiliation{Walther-Mei{\ss}ner-Institut, Bayerische Akademie der Wissenschaften, Garching 85748, Germany}

\author{I.~Tsitsilin}
\affiliation{Technical University of Munich, TUM School of Natural Sciences, Department of Physics, Garching 85748, Germany}
\affiliation{Walther-Mei{\ss}ner-Institut, Bayerische Akademie der Wissenschaften, Garching 85748, Germany}

\author{F.~Wallner}
\affiliation{Technical University of Munich, TUM School of Natural Sciences, Department of Physics, Garching 85748, Germany}
\affiliation{Walther-Mei{\ss}ner-Institut, Bayerische Akademie der Wissenschaften, Garching 85748, Germany}

\author{S.~J.~Glaser}
\affiliation{Technical University of Munich, TUM School of Natural Sciences, Department of Physics, Garching 85748, Germany}
\affiliation{Munich Center for Quantum Science and Technology (MCQST), Schellingstra{\ss}e 4, M\"{u}nchen 80799 , Germany}

\author{M.~Werninghaus}
\email{max.werninghaus@wmi.badw.de}
\affiliation{Technical University of Munich, TUM School of Natural Sciences, Department of Physics, Garching 85748, Germany}
\affiliation{Walther-Mei{\ss}ner-Institut, Bayerische Akademie der Wissenschaften, Garching 85748, Germany}

\author{S.~Filipp}
\affiliation{Technical University of Munich, TUM School of Natural Sciences, Department of Physics, Garching 85748, Germany}
\affiliation{Walther-Mei{\ss}ner-Institut, Bayerische Akademie der Wissenschaften, Garching 85748, Germany}
\affiliation{Munich Center for Quantum Science and Technology (MCQST), Schellingstra{\ss}e 4, M\"{u}nchen 80799 , Germany}

\date{\today}

\begin{abstract}
State-of-the-art single-qubit gates on superconducting transmon qubits can achieve the fidelities required for error-corrected computations.
However, parameter fluctuations due to qubit instabilities, environmental changes, and control inaccuracies make it difficult to maintain this performance.
To mitigate the effects of these parameter variations, we numerically derive gates robust to amplitude and frequency errors using gradient ascent pulse engineering (GRAPE).
We analyze how fluctuations in qubit frequency, drive amplitude, and coherence affect gate performance over time.
The robust pulses suppress coherent errors from drive amplitude drifts over 15 times more than a Gaussian pulse with derivative removal by adiabatic gate (DRAG) corrections.
Furthermore, the robust gates, originally designed to compensate for quasi-static errors, also demonstrate resilience to stochastic, time-dependent noise, which is reflected in the dephasing time.
They suppress added errors during increases in dephasing by up to 1.7 times more than DRAG.
\end{abstract}

\maketitle
\input{Sections/sec1_introduction}
\input{Sections/sec2_robust}
\input{Sections/sec3_drifts}
\input{Sections/sec4_conclusion}
\input{Sections/sec5_acknowledgements}

\bibliography{bib}
\clearpage

\begin{appendices}
    \input{Appendices/appA_setup}
    \input{Appendices/appB_algorithm}
    \input{Appendices/appC_calibration}
\end{appendices}

\end{document}

%% file: Sections/sec1_introduction.tex
\section{Introduction}\label{sec:introduction}
Superconducting qubits have emerged as a leading platform for scalable, error-corrected quantum processors that operate at high speed~\cite{Krinner2022, Google2023, Knill1969}.
In particular, they support fast single-qubit gate operations which can meet the fidelity threshold for scalable quantum error correction (QEC)~\cite{Knill1969, Acharya2025}.
However, maintaining the required level of performance is challenging due to instabilities arising from environmental disturbances and drifts in control hardware that lead to both coherent and stochastic errors~\cite{Sivak2025}.

Coherent errors can arise from persistent parameter offsets, such as crosstalk~\cite{Tripathi2022} and miscalibrations.
They can also occur due to quasi-static drifts that are effectively constant over the duration of an experiment, for example, slow fluctuations in the qubit frequency~\cite{Burnett2019} and instabilities in control electronics~\cite{vanDijk2019}.
These errors are difficult to detect or correct with conventional QEC codes, which are designed primarily to address random Pauli noise~\cite{Kueng2016, Marton2023, Sivak2025}.
In contrast to coherent errors, stochastic errors fluctuate randomly in time, leading to decoherence.
They originate from sources such as two-level systems (TLSs)~\cite{Martinis2005, Agarwal2024}, ionizing radiation~\cite{Li2025}, and quasiparticles~\cite{Riste2013}, and can cause fluctuations in qubit coherence times and frequency~\cite{Muller2015, Burnett2019}.
Changes in the energy relaxation time $T_1$, often linked to interactions with TLSs, occur unpredictably on millisecond to hour timescales~\cite{Carroll2022, Berritta2025} and exhibit spatio-temporal correlations associated with radiation events~\cite{Thorbeck2023, Wilen2021, Gordon2022}.
Similar behaviour is observed for the dephasing time $T_\phi$, where rapid frequency fluctuations from dispersive interactions with TLSs and from quasiparticle tunneling lead to significant reductions in coherence~\cite{Burnett2019, Schlor2019}.
These stochastic errors degrade quantum processor performance and reduce the effectiveness of QEC, often necessitating post-selection to discard data affected by low coherence~\cite{Chen2021, Klimov2024, Acharya2025}.
Robust gate designs that inherently tolerate parameter fluctuations offer a promising path to mitigate both coherent and stochastic errors.

Several methods have been demonstrated for realizing robust gates including open-loop optimization~\cite{Shao2024, Edmunds2020, Carvalho2021, Zhang2025, Huang2017, Le2023}, geometric gates~\cite{Xue2025, Harutyunyan2023, Yi2024,  Amer2025, Dridi2020}, composite pulses~\cite{Tonchev2025, Cummins2003, Shi2024, Kabytayev2014}, sampling-based learning control~\cite{Dong2015, Dong2016, Cykiert2024}, and other advanced optimization methods~\cite{Kuzmanovic2024, Berger2024, Glaser2025, Werninghaus2021, Koch2022}.
Robust gates can be designed to combat fluctuations in various parameters including drive amplitude~\cite{Harutyunyan2023, Cykiert2024}, both drive amplitude and qubit frequency~\cite{Dridi2020, Yi2024, Amer2025, Tonchev2025, Cummins2003, Shi2024, Huang2017, Edmunds2020, Carvalho2021, Le2023, Kuzmanovic2024, Zhang2025, Berger2024}, and qubit-qubit coupling~\cite{Xue2025, Shi2024, Dong2015, Dong2016, Huang2017, Le2023}. 
While robust gates consistently reduce coherent errors in experiments~\cite{Harutyunyan2023, Yi2024, Amer2025, Cykiert2024, Edmunds2020, Carvalho2021, Kuzmanovic2024, Berger2024}, their behaviour in complex noise environments remains less understood~\cite{Koch2022}.
Previous simulations have shown that robust gates, even those designed only for static errors, can offer resilience against stochastic, time-varying noise~\cite{Shao2024, Huang2017, Kabytayev2014}. Indeed, robust gates have been shown to help stabilize operation fidelity over time~\cite{Carvalho2021}; however, a challenge remains to experimentally demonstrate and quantify their robustness under realistic operating conditions.

In this work, we design two robust gates for superconducting transmon qubits through gradient ascent pulse engineering (GRAPE)~\cite{Khaneja2005,Machnes2011}: a Frequency RObust Gate (FROG) and an Amplitude-and-frequency RObust Gate (AROG).
Their robustness landscapes, measured experimentally by adding amplitude and detuning errors to the system, reveal protection in the targeted parameters compared to a standard Gaussian pulse with derivative removal by adiabatic gate (DRAG) corrections~\cite{Motzoi2009, Gambetta2011, Motzoi2013}.
We further investigate operational noise, encompassing both coherent and stochastic errors, discuss potential sources, and analyze their individual contributions to gate error.

%% file: Sections/sec2_robust.tex
\section{Robust gates}\label{sec:robust}
\subsection{Device and Model}
Experiments are carried out on a transmon-type fixed-frequency superconducting qubit~\cite{Koch2007} which has a transition frequency of $\omega_{01}/(2\pi) = 3.9872(2)$~GHz, an anharmonicity $\alpha/(2\pi) = -295.1(2)$~MHz and maximum Rabi rate $\Omega_0/(2\pi)=17.7(5)$~MHz.
The average coherence times are $T_1=45.5 (1)~\mu$s and $T_2^* = 16.81(6)~\mu$s.
The qubit is capacitively coupled to a readout-resonator at $\omega_r/(2\pi) = 6.89041(3)$~GHz with a coupling strength of $g/(2\pi) =72(5)$~MHz.
Throughout this work, the uncertainty of fitted parameters is the expected one-standard-deviation uncertainty, derived from the diagonal elements of the covariance matrix. 
When time series data are available, we report the mean of the measured values, with the uncertainty given by the standard error of the mean.
Further details on the setup are provided in Appendix~\ref{app:setup}.

The dynamics are described by the Hamiltonian
\begin{align}\label{eq:Wright-Hamiltonian}
    \frac{\hat{H}^{\text{R}}}{\hbar}&=\alpha\ket{2}\bra{2}+\delta\mathop{\sum}\limits_{k=1}^2k\ket{k}\bra{k}  \\
    &+\left(1 +\frac{\gamma}{\Omega_0}\right)\left(\frac{\Omega_x(t)}{2}\mathop{\sum}\limits_{k=1}^2{\hat\sigma}_{k,k-1}^x  + \frac{\Omega_{y}(t)}{2}\mathop{\sum}\limits_{k=1}^2{\hat\sigma }_{k,k-1}^y\right) \nonumber
\end{align}
in a frame rotating at the qubit frequency $\omega_{01}$.
In simulations, we include three energy levels $k$ with eigenstates $\ket k$ to capture leakage.
We describe transitions between adjacent energy levels with the operators $\hat\sigma_{k,k-1}^x=\sqrt{k}\left(\ket{k}\bra{k-1}+\ket{k-1}\bra{k}\right)$ and $\hat\sigma_{k,k-1}^y=i\sqrt{k}\left(\ket{k}\bra{k-1} - \ket{k-1}\bra{k}\right)$.
Pulses consist of two control components $\Omega_x(t)$ and $\Omega_y(t)$, which are combined into a single drive signal $\Omega(t) = \Omega_x(t) + i\Omega_y(t)$.
To reflect fluctuations in the system parameters, we have introduced $\delta$ as a frequency error and $\gamma$ as a drive amplitude error in the Hamiltonian.

\subsection{Numerical optimization}
\begin{figure*}[t]
\includegraphics[width=\textwidth]{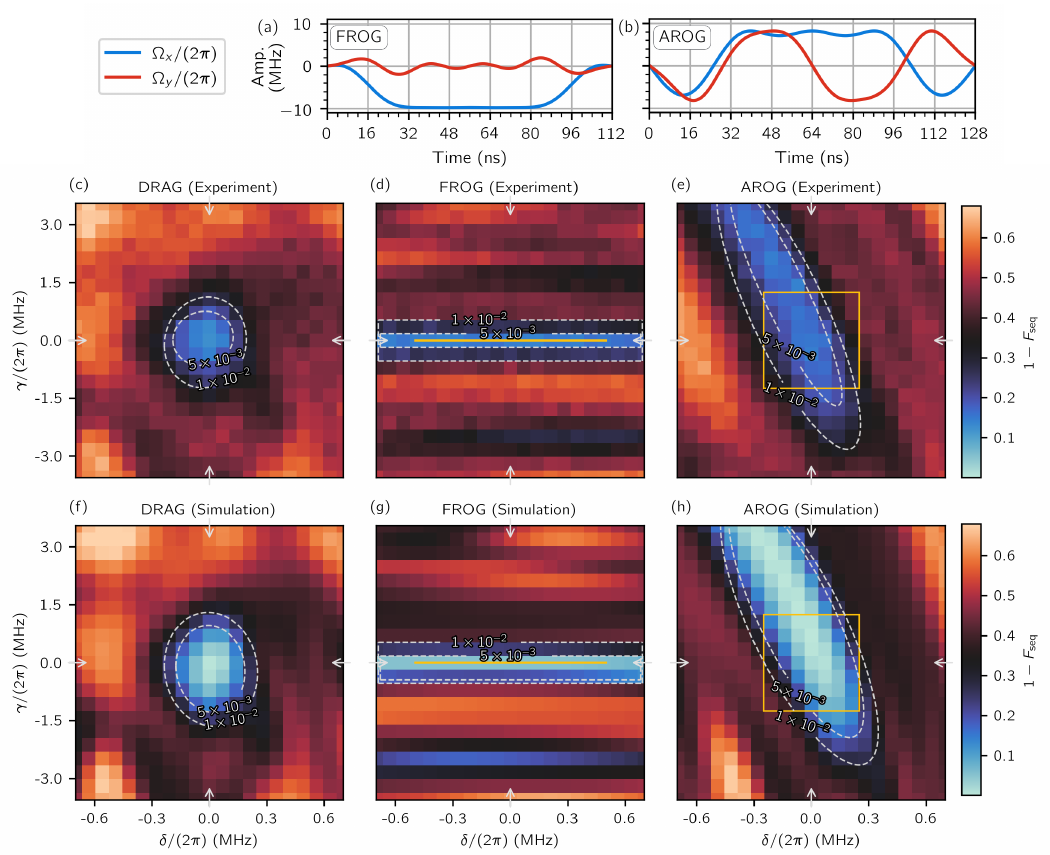}
\caption{\label{fig:landscapes} 
Pulse envelopes for (a) FROG and (b) AROG. The pulses are given by Eq.~\eqref{eq:Wright-FourierPulse}.
Randomized benchmarking sequence error $1 - F_\text{seq}$ as a function of frequency error $\delta$ and amplitude error $\gamma$ for the (c) DRAG, (d) FROG, and (e) AROG gates.
Simulations of the three gates performed using Eq.~\eqref{eq:Wright-Hamiltonian} are shown in (f)-(h) for comparison.
While the colourbar shows sequence error, the contour lines correspond to gate errors $\mathcal{E}_g = 5\times 10^{-3}$ and $\mathcal{E}_g = 1\times 10^{-2}$, estimated from the sequence error using Eq.~\eqref{eq:Wright-ORBIT}.
The yellow line and box indicate the range of errors over which FROG and AROG are optimized.
The arrows indicate linecuts at zero amplitude error and at zero frequency error, with the corresponding line plots shown in Fig.~\ref{fig:linecuts}.}
\end{figure*}
To compensate for frequency and amplitude errors, we derive robust gates using numerical optimization.
To force the pulse to start and end at zero, ensure symmetries, and keep pulse variations slow, we choose a pulse ansatz in the Fourier basis of the form
\begin{equation}\label{eq:Wright-FourierPulse}
\begin{aligned}
    &\Omega_x(t) = \Omega_0\sum_{n=1}^5 a_n \sin((2n - 1)\pi t/t_g)~, \\
    &\Omega_y(t) = \Omega_0\sum_{n=1}^5 b_n \sin(2n\pi t/t_g)~,
\end{aligned}
\end{equation}
where $a_n$, $b_n$ represent Fourier series coefficients, $t_g$ is the gate length, and $n$ enumerates the Fourier components.
In all cases, we optimize ten coefficients, five each for $a_n$ and $b_n$, to shape the pulse defined in Eq.~\eqref{eq:Wright-FourierPulse}. 
This choice promotes smooth pulse profiles, avoiding abrupt changes or high-frequency components, which are expected to be less sensitive to pulse distortions and other undesirable effects~\cite{Hyyppa2024}.

We optimize the Fourier series coefficients in Eq.~\eqref{eq:Wright-FourierPulse} using the GRAPE algorithm adapted to analytically shaped functions~\cite{Skinner2010}.
GRAPE works by discretizing the time evolution under the parametrized pulse shape.
The fidelity of the resulting operation is evaluated by a cost function, and gradients of this cost function with respect to the pulse parameters are computed efficiently using forward and backward propagation.
The parameters are then updated iteratively along the gradient direction to increase fidelity.

To achieve robustness, we optimize over a discrete set of Hamiltonians, as defined in Eq.~\eqref{eq:Wright-Hamiltonian}, which differ by variations in the detuning $\delta$ and amplitude $\gamma$ scaling parameters. 
Specifically, we consider deviations in \(\delta \in [\delta_{\text{min}}, \delta_{\text{max}}]\) and \(\gamma \in [\gamma_{\text{min}}, \gamma_{\text{max}}]\), sampling a finite set of values across these intervals.
For each sampled pair \((\delta_i, \gamma_j)\), we define a corresponding perturbed Hamiltonian \(H_{ij}(t)\) and simulate the system evolution.
For each pair of parameters \((\delta_i, \gamma_j)\), we define the gate error as:
\begin{equation}\label{eq:Wright-GateFidelity}
    \mathcal{E}_g^{ij} = 1- F_g^{ij} = 1 - \frac14\left|\sum_{k=0}^1 \bra{k}U_\text{T}^\dagger U_{ij}(t_g)\ket{k}\right|^2,
\end{equation}
where \(U_{ij}(t_g)\) is the time-evolution operator at time \(t_g\) solving $\dot{U}_{ij}(t)=-H_{ij}(t)U_{ij}(t)$ and $U_\text{T}$ is the target unitary.
As a cost function we use the gate error averaged over the entire parameter set:
\begin{equation}\label{eq:GrapeCostFun2}
J = \frac{1}{N_{\delta} N_{\gamma}} \sum_{i=1}^{N_{\delta}} \sum_{j=1}^{N_{\gamma}} \mathcal{E}_g^{ij},
\end{equation}
where \(N_{\delta}\) and \(N_{\gamma}\) are the number of sampled values of \(\delta\) and \(\gamma\), respectively.
This ensemble-averaged cost ensures that the optimized pulse performs well not just for the nominal Hamiltonian, but also across the expected range of deviations.
By iteratively updating the pulse shape using the gradient of the average cost with respect to the control parameters, the GRAPE algorithm converges to a solution that is robust to fluctuations in \(\delta\) and \(\gamma\).

We target $U_T = X_{\pi/2}$ as it can be used to form a basis for single-qubit operations $\{X_{\pi/2}, Z_\theta\}$, with $Z_\theta$ implemented as virtual $Z$ gates~\cite{McKay2017} for an arbitrary angle $\theta$.
We use the same pulse ansatz from Eq.~\eqref{eq:Wright-FourierPulse} to obtain gates protected against different parameter deviations by adjusting the range of error values used in the cost function.
We derive the FROG gate by setting $\gamma_\text{min} = \gamma_\text{max} = 0$ and $\delta/(2\pi) \in [-0.5, 0.5]$~MHz with $t_g = 112$~ns, and the AROG gate by choosing and $\gamma/(2\pi) \in [-1.2, 1.2]$~MHz and $\delta/(2\pi) \in [-0.25, 0.25]$~MHz with $t_g = 128$~ns. 
The pulses envelopes for FROG and AROG are shown in Fig.~\ref{fig:landscapes}~(a) and (b), respectively.
Further details about the optimization method are provided in Appendix~\ref{app:algorithm}.
The open-source code is also available online~\cite{PythonGRAPE_SQC2025}.

\subsection{Experimental validation}
On the device, we first measure the performance of the gates with no added error using randomized benchmarking (RB) experiments~\cite{Knill2008, Magesan2011, Wood2018}.
To provide a baseline for comparison, we use a Gaussian pulse with DRAG corrections of duration 128 ns, chosen to ensure comparable exposure to noise relative to the robust gates.
The error for all gates is comparable at $\mathcal{E}_\text{DRAG}=1.80(4)\times 10^{-3}$, $\mathcal{E}_\text{FROG} = 1.70(4)\times 10^{-3}$, and $\mathcal{E}_\text{AROG} = 1.94(3)\times 10^{-3}$.
Details on the calibration to achieve these gate errors are presented in Appendix~\ref{app:cal}.

To determine the gate error as a function of parameter deviations, we introduce static parameter offsets in a range $\delta/(2\pi) \in [-0.7, 0.7]$~MHz and $\gamma/(2\pi) \in [-3.5, 3.5]$~MHz.
The parameter offsets are chosen such that they span around three times the respective ranges over which AROG is optimized and are sampled on a $21\times21$ grid.
To efficiently explore a large parameter space, we take the RB sequence error at a single length $N_C=60$, averaged over 10 randomizations, as a proxy for the gate error~\cite{Kelly2014}.
The average gate fidelity $F_g$ is approximately related to the average sequence fidelity $F_\text{seq}$ at a single length $N_C$ as
\begin{equation}\label{eq:Wright-ORBIT}
    \begin{aligned}
        F_\text{seq} \approx A \left(F_g\right)^{N_C N_g} + B
    \end{aligned}
\end{equation}
where $N_g$ refers to the average number of physical gates required to implement each Clifford gate and $A, B$ take into account state preparation and measurement (SPAM) errors~\cite{Magesan2012}.
For our chosen gate decomposition $\{X_{\pi/2}, -X_{\pi/2}, Y_{\pi/2}, -Y_{\pi/2}\}$, we have $N_g=1.25$, where $Y$-rotations are obtained by applying the same pulse envelope as for $X$, but with the drive phase shifted by $\pm \pi/2$ relative to the $X$-axis.
In full RB experiments, the parameters $A$ and $B$ are found by the fit to data.
We work backwards from the sequence fidelity, setting $A= 0.435$ and $B = 0.507$ according to our readout characterization matrix, assuming similar behaviour of the sequence fidelity $F_\text{seq}$ as in the full RB measurements.

As the DRAG gate is not designed to compensate for parameter fluctuations, the sequence error $1 - F_\text{seq}$ increases with amplitude errors $\gamma$ and frequency errors $\delta$, as shown in Fig.~\ref{fig:landscapes}~(c). 
The FROG gate expands the protected region along the detuning dimension but compresses it along the amplitude dimension, while the AROG gate adds amplitude protection without sacrificing robustness to static frequency noise.
In experiment, the FROG gate maintains a sequence error $1-F_\text{seq}$ corresponding to a gate error $\mathcal{E}_g \leq 10^{-2}$ (estimated from Eq.~\eqref{eq:Wright-ORBIT}) over 100\% of the range targeted by its optimization, as shown in Fig.~\ref{fig:landscapes}~(d), compared to 37.5\% for the DRAG and AROG gates. 
We adopt $\mathcal{E}_g = 10^{-2}$ as a threshold gate error throughout and note its relevance for QEC thresholds~\cite{Kueng2016}.
Over the rectangular region of parameter variations targeted in the AROG optimization, indicated by the yellow box in Fig.~\ref{fig:landscapes}~(e), the AROG gate has $\mathcal{E}_g \leq 10^{-2}$ over 67.2\% of the area, an improvement over 46.9\% for the reference DRAG gate and 35.9\% for the FROG gate.
Both FROG and AROG also have low error $\mathcal{E}_g \leq 10^{-2}$ outside the targeted parameter ranges.
For the FROG gate, the protection is simply extended along the frequency error axis, while the low error area for the AROG gate forms a tilted ellipse.
This shape is an artifact of the optimization, and we address its implications further in the discussion.
The experimental results are supported by simulations using the Hamiltonian in Eq.~\eqref{eq:Wright-Hamiltonian}, with good agreement shown in Fig.~\ref{fig:landscapes}(f)-(h).
All gates exhibit an increase in error of approximately $5\times10^{-3}$ in experiment compared to simulation, due to decoherence and other noise not included in the model.

\begin{figure}[t]
\includegraphics[width=0.48\textwidth]{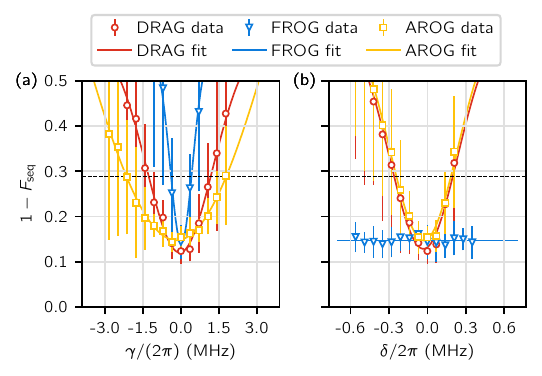}
\caption{\label{fig:linecuts} 
Randomized benchmarking sequence error $1 - F_\text{seq}$ for (a) amplitude errors $\gamma$ and (b) frequency errors $\delta$ for the DRAG, FROG, and AROG gates.
Each panel shows a line cut from the 2D plots shown in Fig.~\ref{fig:landscapes}, along the directions indicated by the grey arrows.
The error bars reflect the standard deviation across different randomizations
The data are fitted with Gaussian functions.
The horizontal line indicates a gate error $\mathcal{E}_g = 10^{-2}$, estimated from the sequence fidelity using Eq.~\eqref{eq:Wright-ORBIT}.}
\end{figure}

To further quantify the robustness of the gates, we fit a Gaussian to two linecuts, shown in Fig.~\ref{fig:linecuts}, from the sequence error landscape.
The linecuts are at zero frequency error and at zero amplitude error, for each gate, as indicated by the gray arrows in Fig.~\ref{fig:landscapes}.
The AROG gate adds amplitude protection, with $\mathcal{E}_g \leq 10^{-2}$ over a range $\gamma/(2\pi) \in [-2.2, 1.8]$~MHz, compared to $[-1.2, 1.2]$~MHz for DRAG and  $[-0.4, 0.4]$~MHz for FROG, as shown in Fig.~\ref{fig:linecuts}~(a). 
The amplitude protection does not compromise robustness to static frequency noise for the AROG gate, which remains comparable to that of a DRAG gate; both achieve $\mathcal{E}_g \leq 10^{-2}$ over $\delta/(2\pi) = \pm 0.2$~MHz.
The FROG gate has sequence error corresponding to $\mathcal{E}_g \leq 10^{-2}$ over the full measured frequency range $\delta/(2\pi) = \pm 0.7$~MHz, as seen in Fig.~\ref{fig:linecuts}~(b).  

%% file: Sections/sec3_drifts.tex
\section{Parameter fluctuations over time}\label{sec:drifts}
To understand when robustness is beneficial in realistic noise environments, we further investigate gate performance as the parameters fluctuate over time.
We track amplitude and frequency errors as well as qubit coherence times over 11 days.
We measure amplitude deviations $\gamma$ with an error amplification sequence $X_{\pi/2}^{(2n + 1)}$. The final population fitted to a cosine as a function of the number of repetitions $n$~\cite{Sheldon2016}, described in detail in Appendix\,\ref{app:cal}.
We measure the qubit frequency drift $\delta$ and dephasing time $T_2^*$ from Ramsey experiments for various delays $\tau_r$.
The $T_1$ time is extracted from a decay experiment where the qubit is excited and the remaining population is measured after delays $\tau_d$.
The pure dephasing time $T_\phi$ can be calculated from these measurements using the relation $\frac{1}{T_\phi} = \frac{1}{T_2^*} - \frac{1}{2T_1}$.
To ensure the measured quantities reflect the same noise environment on average, the experiments are interleaved on a shot-by-shot basis:
a new set $(n, \tau_r, \tau_d)$ of amplitude error repetitions $n$, Ramsey delay time $\tau_r$, and $T_1$ delay time $\tau_d$ is measured in each experiment (each shot). 
In this manner, we simultaneously iterate over $n \in \{0, \ldots, 34\}$,  $\tau_r \in [0, 40]\,\mu\text{s}$ and $\tau_d \in [1\,\text{ns}, 100\,\mu\text{s}]$, each discretized into 35 points.
Having iterated through all 35 sets of $(n, \tau_r, \tau_d)$, we repeat this sequence $2^{12}$ times.
The full characterization measurement is repeated 10 times, which takes approximately 10 minutes.

To determine how the parameter fluctuations affect gate error, we perform RB experiments between the full characterization measurements. 
The DRAG, FROG, and AROG gate performance is measured by interleaving data points across randomizations, ensuring that each gate is sampled uniformly over the parameter drifts.
Each RB experiment contains $N_r = 60$ randomizations and $N_s = 16$ sequence lengths for each gate.
Averaging over $2^8$ repetitions, the full measurement runs for approximately 30 minutes.
The cycle of characterization and RB is repeated continuously for 10 hours per day over 11 consecutive days.

\begin{figure}
    \centering
    \includegraphics[width=0.48\textwidth]{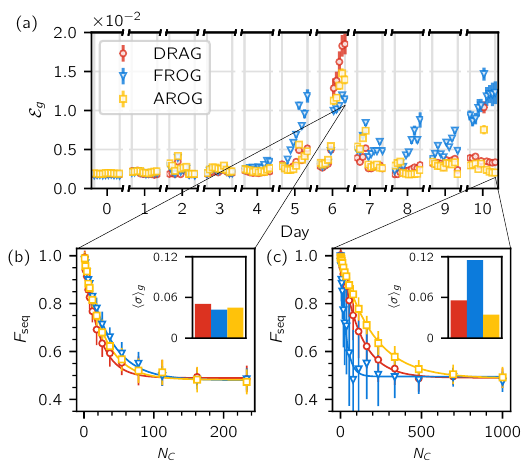}
    \caption{(a) Gate error $\mathcal{E}_g$ for the DRAG, FROG, and AROG gates extracted from RB experiments for 10 hours each day for 11 days. RB curves for (b) day 6 and (c) day 10 are connected by lines to time points of interest in (a). The error bars at each sequence length $N_C$ reflect the standard deviation across different randomizations, and the insets show the mean standard deviation $\langle \sigma \rangle_g$ over all sequence lengths.}
    \label{fig:RBSeries}
\end{figure}

The gate error $\mathcal{E}_g$ stays relatively stable for all gates in the first 5 days, but fluctuates thereafter, with increases by as much as $1.7(1) \times 10^{-2}$ on day 6 and $1.30(8) \times 10^{-2}$ on day 10, as shown in Fig.\,\ref{fig:RBSeries}\,(a).
The different pulses exhibit varying levels of robustness on these days, as shown by their RB performance depicted in Fig.\,\ref{fig:RBSeries}\,(b) and (c), respectively.
In the following, we examine the parameter fluctuations in detail and analyze how they contribute to the observed variations in gate error, with particular attention to day 6, where the qubit experienced strong dephasing, and day 10, where large amplitude errors occurred.

\subsection{Amplitude Drifts}
\begin{figure}[t]
\includegraphics[width=0.48\textwidth]{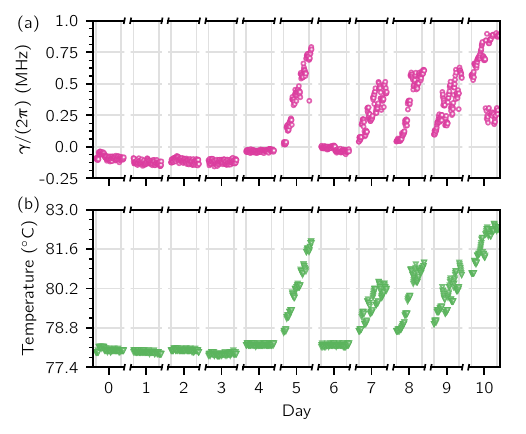}
\caption{\label{fig:ampSeries} 
(a) Amplitude error $\gamma$ for an $X_{\pi/2}$ gate extracted from error amplification experiments and (b) internal temperature for the device used to generate the drive signals,  measured independently for 10 hours each day for 11 days.
}
\end{figure}

Over the 11 days of measurement, we observe gradual drive amplitude drifts by over $\gamma/(2\pi) = 0.90$\,MHz (6\%) from the initial calibrated value, with the slow drift accompanied by smaller, short-term fluctuations.
These drifts occur on specific measurement days (day 5 and days 7-10), whereas on other days (days 0-4 and day 6) the amplitude deviations remain almost constant near zero, as shown in Fig.\,\ref{fig:ampSeries}.
We attribute amplitude drifts mainly to temperature instabilities on the control device.
The amplitude deviations have a Pearson correlation of $r = 0.94 \quad \text{(95\% CI: 0.933 – 0.946)}$ with the internal temperature of the device used to generate the drive signal, shown in Fig.\,\ref{fig:ampSeries}\,(b).
The effects of these drifts are evident in the changes in gate error $\mathcal{E}_g$ over time.
The FROG gate, which has increased sensitivity to amplitude deviations, experiences increased gate errors caused by the drifts on days 5 and 7-10, as shown in Fig.\,\ref{fig:RBSeries}\,(a).

Although less affected by these events, the DRAG and AROG gates still exhibit increased error.
Day 10 provides a clear example of this effect: the amplitude deviates by $\gamma/(2\pi) = 0.88$\,MHz as the device heats up by nearly $5\degree$C. Consequently, the DRAG gate error increases by 87\% from the initial calibration with $\Delta\mathcal{E}_\text{DRAG} = 1.57(9)\times 10^{-3}$.
In contrast, the AROG gate error increases by only 6\% with $\Delta\mathcal{E}_\text{AROG} = 1.1(5)\times 10^{-4}$, corresponding to a 15-fold improvement over DRAG in added error, as shown in Fig.\,\ref{fig:RBSeries}\,(c). 
The effect of amplitude errors is also captured by the spread of the RB, which increases in the presence of coherent errors~\cite{Ball2016}.
Before the amplitude drift, all gates show a mean standard deviation over sequence lengths of $\langle \sigma \rangle_g = 0.03$.
During the drift, the spread increases substantially for the FROG gate, reaching $\langle \sigma \rangle_{\text{FROG}} = 0.11$, and it doubles for the DRAG gate to $\langle \sigma \rangle_{\text{DRAG}} = 0.06$, whereas the AROG gate maintains the same value of $\langle \sigma \rangle_{\text{AROG}} = 0.03$, as shown in the inset of Fig.\,\ref{fig:RBSeries}\,(c).

\subsection{Frequency Drifts}
\begin{figure}[t]
\includegraphics[width=0.48\textwidth]{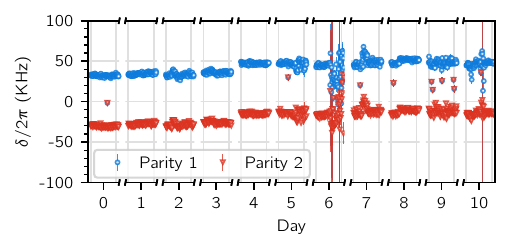}
\caption{\label{fig:freqSeries} 
(a) Qubit frequency extracted from Ramsey experiments for 10 hours each day in an 11 day period. The two distinct frequency regions indicate the even (blue circle) and odd (red triangle) parity states of the qubit.
}
\end{figure}

We observe qubit frequency drifts by a maximum of $\delta = 77(3)$\,kHz in the eleven day period, well within the protected area for all three gates, as shown in Fig.\,\ref{fig:freqSeries}.
The qubit exhibits two distinct frequencies which can be attributed to quasi-particle induced charge parity switching~\cite{Riste2013}.
The difference between frequencies remains stable around $61.4(4)$\,kHz. 
Compared to the effect of other parameters, the small quasi-static drifts in qubit frequency do not compromise gate performance.
The frequency measurement is imprecise on day 6 and at some points on day 10 due to low qubit coherence, which we discuss in the next section.

\subsection{Coherence Time Drifts}
\begin{figure}[t]
\includegraphics[width=0.48\textwidth]{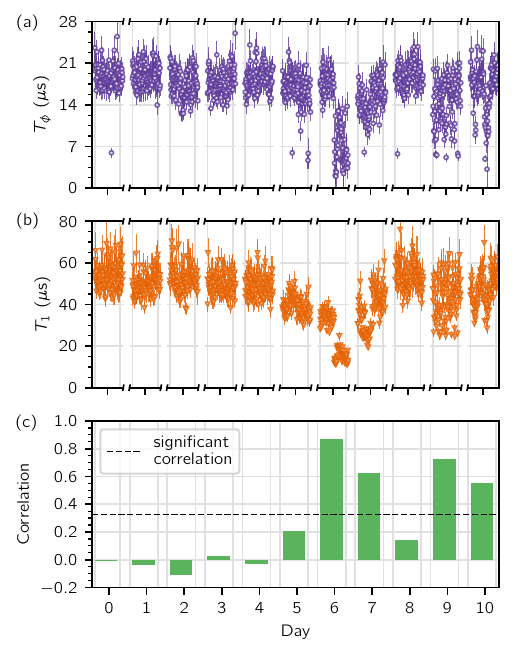}
\caption{\label{fig:coherenceSeries} 
(a) Qubit pure dephasing time $T_\phi$ and (b) qubit decay time $T_1$ extracted from Ramsey and decay experiments interleaved on a shot by shot basis for 10 hours each day in an 11 day period. (c) Pearson correlation between $T_\phi$ and $T_1$ each day. The dashed horizontal line indicates a correlation of 0.325 (p-score 0.001).}
\end{figure}

The pure dephasing time $T_\phi$ and decay time $T_1$ vary from 1\,$\mu$s to 26\,$\mu$s and 11\,$\mu$s to 80\,$\mu$s, respectively.
The fluctuations are largest on days 6–7 and 9–10, as shown in Fig.\,\ref{fig:coherenceSeries}.
On these days, the two time series have a Pearson correlation greater than $r=0.325$, which we choose as a threshold for meaningful correlation because it corresponds to a statistically significant p-value of 0.001 for our measurements of 100 data points per day.
The correlations suggest a common source of noise affecting both relaxation and dephasing.
The $T_\phi$ time reflects random fast frequency fluctuations beyond the quasi-static noise captured by the central frequency drift, and we observe resilience to this stochastic noise by the robust gates, particularly FROG.

The lowest qubit coherence occurs on day 6, with reduced $T_1$ and $T_\phi$ for a period of around 6 hours.
The gates experience an increase in error of $\Delta\mathcal{E}_\text{FROG} = 1.02(3)\times 10^{-2}$, $\Delta\mathcal{E}_\text{AROG} = 1.28(6) \times 10^{-2}$ and $\Delta\mathcal{E}_\text{DRAG}= 1.7(1)\times 10^{-2}$ from the initial calibrated values.
Operating with a FROG gate leads to a 1.7-fold reduction in additional error, and with AROG a 1.3-fold reduction, during the period of low qubit coherence, as shown by the RB curves in Fig.\,\ref{fig:RBSeries}\,(b).
During another reduction in $T_\phi$ on day 10, which lasts around an hour, the DRAG gate error again rises above $10^{-2}$ by an increase of $\Delta\mathcal{E}_\text{DRAG} = 8.6(8)\times 10^{-3}$ over the day 0 value.
This time the AROG gate shows an increase of $\Delta\mathcal{E}_\text{AROG} = 5.6(6)\times 10^{-3}$, 1.5 times less additional error than the DRAG gate, while we observe an increase in the FROG gate error by $\Delta\mathcal{E}_\text{FROG} = 1.80(8)\times 10^{-2}$, mainly due to a simultaneous amplitude drift of $\gamma/(2\pi) = 0.83$\,MHz.
A more detailed analysis is required to disentangle the individual contributions of the parameter fluctuations to the observed gate-error.

\subsection{Sensitivity Analysis}
To quantify the change in gate error per parameter, we perform a Ridge regression on the time series data.
The regression evaluates the sensitivity $\Delta\mathcal{E}_g/\Delta x$ with respect to $x = \gamma, \Gamma_\phi, \Gamma_1$ for each gate $g =$\,DRAG, FROG, AROG.
We use the rates $\Gamma_\phi= 1/T_\phi, \Gamma_1 = 1/T_1$ rather than coherence times to capture their nonlinear relationship with gate error and we omit qubit frequency drifts $\delta$, which have negligible effects in our experiments.
At each time point, we have a vector $\mathbf{x}_k = (\gamma^k, \Gamma_\phi^k, \Gamma_1^k) \in \mathbb{R}^3$ of inputs, which the regression maps to an output $\boldsymbol{\mathcal{E}}_k = (\mathcal{E}^k_\text{DRAG}, \mathcal{E}^k_\text{FROG}, \mathcal{E}^k_\text{AROG}) \in \mathbb{R}^3$ consisting of the error for each gate.
Before fitting, we normalize the parameters as $\tilde{\mathbf{x}}_k = (\mathbf{x}_k - \boldsymbol{\mu}) / \boldsymbol{\sigma}$ using the mean $\boldsymbol{\mu} = (\mu_\gamma, \mu_{\Gamma_\phi}, \mu_{\Gamma_1})$ and standard deviation $\boldsymbol{\sigma} = (\sigma_\gamma, \sigma_{\Gamma_\phi}, \sigma_{\Gamma_1})$ of each parameter over time.
The Ridge regression minimizes the mean squared error
\begin{equation}\label{eq:Wright-MSE}
 \frac{1}{m} \sum_{k=1}^m \left(\boldsymbol{\mathcal{E}}_k - \tilde{\mathbf{x}}_k^\top \tilde{\mathbf{w}} \right)^2 
+ \boldsymbol{\lambda}\lVert \tilde{\mathbf{w}} \rVert_2^2
\end{equation}
where $m=110$ is the number of samples and $\tilde{\mathbf{w}} \in \mathbb{R}^9$ is the weight vector to be fitted which quantifies the change in error per change in normalized parameter $\tilde{w}_{x,g} = \Delta\mathcal{E}_g/\Delta\tilde{x}$.
To obtain the coefficients in physical units, the regression weights $\tilde{\mathbf{w}}$ (fit on normalized features $\tilde{\mathbf{x}}$) are rescaled by the corresponding normalization factors to give the sensitivities $w_{x, g} = \tilde{w}_{x,g}/\sigma_x = \Delta\mathcal{E}_g/\Delta x$ for each parameter $x$ and gate $g$.
The final fit for all gates has an $R^2$ value of 0.73, where we impose $\boldsymbol{\lambda} \geq 0$ for regularization to avoid overfitting and to keep the sensitivities within a physically meaningful range.
The sensitivities in physical units are listed in Tab.\,\ref{tab:RidgeSensitivities}.
\begin{table}[h]
\centering
\begin{tabular}{|c|c|c|c|}
\hline
& $w_\gamma$ & $w_{\Gamma_\phi}$ & $w_{\Gamma_1}$ \\
     & ($10^{-2}$ MHz$^{-1}$) & ($10^{-2}$ MHz$^{-1}$) & ($10^{-2}$ MHz$^{-1}$)  \\
\hline
DRAG  & $0.08$ & $8.11$ & $4.28$ \\
FROG  & $0.80$ & $4.72$ & $3.73$ \\
AROG & $\approx 0$ & $6.15$ & $4.27$ \\
\hline
\end{tabular}
\caption{Scaled weights from a Ridge regression model for three parameters: amplitude deviation $\gamma$, dephasing rate $\Gamma_\phi$, and relaxation rate $\Gamma_1$. The values represent the change in gate error per parameter $w_x = \Delta\mathcal{E}/\Delta x$ and are fit for DRAG, FROG, and AROG gates.}
\label{tab:RidgeSensitivities}
\end{table}

To verify the physical relevance of the fit, we compare the sensitivity to decay $w_{\Gamma_1}$ to an analytical formula for coherence limits.
From the relationship $\mathcal{E}_g \approx \frac{t_g}{3}\Gamma_1$~\cite{Wood2018}, we calculate $w_{\Gamma_1}^{\text{analytical}} = 4.27 \times 10^{-2}$\,MHz$^{-1}$ for DRAG and AROG and $w_{\Gamma_1}^{\text{analytical}} = 3.73 \times 10^{-2}$\,MHz$^{-1}$ for FROG, a near match in all cases.
As expected from our design and initial experiments, the FROG gate shows a pronounced sensitivity to amplitude errors, while the DRAG gate is an order of magnitude less sensitive and the AROG gate exhibits no dependence within the range of the observed drifts.
As well, the FROG and AROG gates are less sensitive to dephasing than the DRAG gate, with the FROG gate about 1.7 times and the AROG gate about 1.3 times more robust, illustrating how  robustness to static frequency errors can also add protection during fluctuations in $T_\phi$.

%% file: Sections/sec4_conclusion.tex
\section{Discussion and Conclusion}\label{sec:discussion}
We have demonstrated the effectiveness of robust single-qubit gates in maintaining stable gate operations, both during quasi-static parameter mismatches and stochastic time-dependent errors.
Using a gradient-based numerical optimization method, we generated a Frequency RObust Gate (FROG) that has protection against detuning errors and an Amplitude-and-frequency RObust Gate (AROG) that is robust against amplitude errors without sacrificing protection to frequency errors.
Under controlled static variations of detuning and amplitude, the FROG and AROG pulses maintain gate errors below $10^{-2}$ for parameter deviations up to three times larger than those tolerated by DRAG. 
The convergence of the RB measurements indicates no significant leakage compared to DRAG, which is achieved by including the third level in the optimization Hamiltonian. 

We further relate fluctuations in drive amplitude, qubit frequency, and coherence times to the gate error of the different pulse designs over time.
During the largest amplitude drift on day 10, the AROG gate experienced over 15 times less added error than the DRAG gate.
Additionally, RB revealed a narrower distribution of sequence fidelity for the AROG gate compared to the DRAG gate during this drift. 
The broader spread across randomizations in the DRAG gate benchmarking data indicates larger coherent errors which lead to a higher worst-case error rate~\cite{Ball2016}.
This is particularly relevant for fault-tolerant quantum computing, where the worst-case error plays a critical role in determining whether the threshold of QEC codes will be met~\cite{Knill1969, Kueng2016, Marton2023}.
We trace back the amplitude deviations to temperature variations of the control device, with fluctuations as large as 1\% per degree celcius.
Temperature changes in the lab, heating of other equipment, and internal heating of the device can all contribute.
When errors in classical control hardware cause drifts, a processor using AROG gates can continue to operate without re-calibration.

Although the gates are optimized to suppress coherent errors, previous simulations suggest that robust designs could also provide resilience to stochastic errors such as fast frequency fluctuations from TLSs~\cite{Shao2024, Huang2017, Kabytayev2014}.
We experimentally confirm this effect for the FROG and AROG gates, which experienced smaller fidelity losses than DRAG during periods of high dephasing, with up to 1.7 times and 1.5 times less added gate error respectively.
The reductions in $T_\phi$ were correlated with reductions in $T_1$ and occurred on minute to hour time scales, which is consistent with prior reports on TLS-induced noise signatures~\cite{Muller2015, Fowler2018, Carroll2022, Burnett2019, Schlor2019, Thorbeck2023, Wilen2021}.
The robustness has practical implications for the operation of superconducting processors.
Advances in shielding, design, and fabrication have reduced the density and coupling of TLS defects~\cite{Place2021, Wang2022, Deng2023, Bal2024, Ganjam2024, Place2021, Wang2022, Deng2023, Bal2024, Ganjam2024, Mergenthaler2021, Mergenthaler2021_2, Dial2016, Wang2015} and in-situ control strategies to actively suppress TLS noise have also been proposed~\cite{Bilmes2020, Bilmes2021, Lisenfeld2023, Zhao2022, Kim2025, Chen2025}, but strongly coupled TLSs and spatio-temporal correlations remain a major error source~\cite{Steffen2017, Mohseni2025}.
In the presence of stochastic frequency-related noise, including from TLSs, which are difficult to avoid, control, or mitigate in other ways, robust gates may provide sufficient resilience to preserve computational usefulness.

Future work involves building on our understanding of gate performance and noise effects to further harness the potential of the gradient-based optimization method and the resulting robust gates introduced here.
Both FROG and AROG exhibit regions of low error extending beyond the parameter ranges explicitly defined in their optimization cost functions, as can be seen in Fig.\,\ref{fig:landscapes}.
For the FROG gate, this low-error extension follows the targeted frequency-error axis, while for AROG it forms a tilted elliptical region in the amplitude–frequency plane, indicating a dependence of the two principle components.
Such a dependence is advantageous only when amplitude and frequency errors are correlated, which we do not observe in our experiments.
Alternative cost functions could yield pulse designs with improved robustness characteristics.
For example, incorporating weighted averages over the target parameter ranges or explicitly accounting for cross-sensitivities in the cost function could suppress unwanted tilting and concentrate robustness where it is most beneficial.
Moreover, the same optimization framework could be extended to explicitly target stochastic noise processes.
By integrating models of fast frequency fluctuations into the optimization on top of the static variations used in this work, subsequent pulse designs could further enhance the protection against stochastic noise already demonstrated by the FROG and AROG gates.
Future implementations could combine robust pulse designs with classical stabilization of control hardware, tailoring the balance between frequency and amplitude robustness to each system’s dominant noise sources.
This hybrid strategy would maximize performance while minimizing calibration overhead.

%% file: Sections/sec5_acknowledgements.tex
\section{Acknowledgements}\label{sec:acknowledgement}
We acknowledge financial support from the German
Federal Ministry of Education and Research via the funding program \textit{Quantum Technologies -- From Basic Research to the Market} under contract number 13N15680 \textit{GeQCoS} and 13N16188 \textit{MUNIQC-SC}; the Deutsche Forschungsgemeinschaft (DFG, German Research Foundation) via project number FI2549/1-1 and Germany’s Excellence Strategy EXC-2111-390814868 \textit{MCQST}; and the European Union’s Horizon research and innovation program through the projects \textit{OpenSuperQPlus100} (Grant-Nr. 955479) and \textit{MOlecular Quantum Simulations (MOQS)} (Grant-Nr. 955479).
We also acknowledge support from the the EU MSCA Cofund I\textit{nternational, Interdisciplinary, and Intersectoral Doctoral Program in Quantum Science and Technologies (QUSTEC)} (Grant-Nr. 847471).
This research is part of the \textit{Munich Quantum Valley}, supported by the Bavarian state government with funds from the Hightech Agenda Bayern Plus.
Emily Wright acknowledges support from the \textit{International Max Planck Research School for Quantum Science and Technology (IMPRS-QST)}.

%% file: Appendices/appA_setup.tex
\section{\texorpdfstring{\\* \vspace{2mm}}~Experimental setup}\label{app:setup}
\begin{figure}[t]
\includegraphics[width=0.48\textwidth]{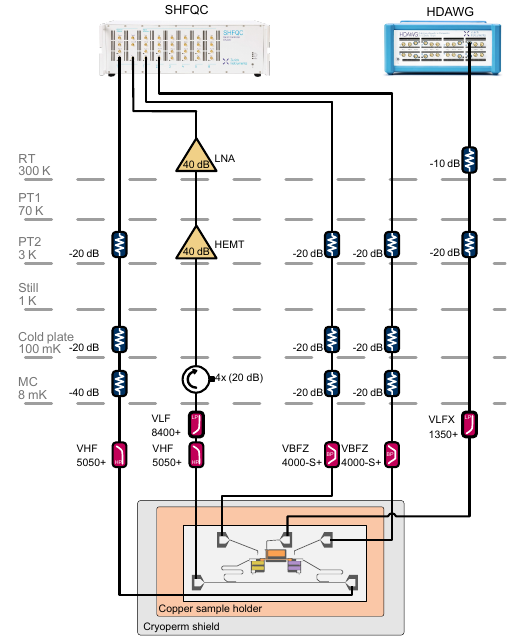}
\caption{\label{fig:setup} 
Experimental setup and schematic of the wiring for the superconducting qubit experiment. The qubits are highlighted in yellow and purple, and the tunable coupler is colored orange in the schematic of the chip.
}
\end{figure}

Our experiments are performed on one qubit from a pair of transmon-type fixed-frequency superconducting qubits~\cite{Koch2007} coupled by a flux tunable coupler~\cite{McKay2016, Roth2017}.
The qubits are controlled and measured using a \textit{Super High Frequency Qubit Controller} (SHFQC) from Zurich Instruments.
A schematic of the experimental setup is shown in Fig.~\ref{fig:setup}.
We use charge-coupled direct drive lines on the chip for qubit control.
The readout resonators are coupled to a common feedline.
The feedline is attenuated by -80 dB distributed over the temperature stages and filtered with a 5.5 GHz high-pass filter (VHF-5050+).
The drive lines are attenuated by -60 dB distributed over the different temperature stages and filtered with a 3.5 GHz to 4.5 GHz band pass filter (VBFZ-4000-S+).
We use two 4-12 GHz isolators on the output line.
The output signal is amplified by a 4-8 GHz 40 dB HEMT cryogenic low-noise amplifier (LNF-LNC4 8G) thermalized at the 3 K stage.
The signal is amplified by a 40 dB low-noise room temperature amplifier (DBLNA104000800A), before being digitized at the SHFQC input.
The flux is controlled using a Zurich Instruments \textit{High-density, multi-channel Arbitrary Waveform Generator} (HDAWG).
The flux line is attenuated at room temperature by -10 dB and by -20 dB at the 3 K stage and filtered using a 1.35 GHz low-pass filter (VLFX-1350+).
We select the qubit with higher coherence times and we bias the tunable coupler at its sweet spot.

%% file: Appendices/appB_algorithm.tex
\section{\texorpdfstring{\\* \vspace{2mm}}~Optimization Algorithm}\label{app:algorithm}
\paragraph*{Cost function.}
We use the cost function described in Eq.~\eqref{eq:GrapeCostFun2} of the main text.
To prevent excessively large pulse amplitudes, we impose constraints of the form: 
\begin{equation}
    \begin{aligned}
        & -\eta\Omega_0\leq\Omega_x\leq\eta\Omega_0,\\
        & -\eta\Omega_0\leq\Omega_y\leq\eta\Omega_0,
    \end{aligned}
    \label{eq:PulseConstr}
\end{equation}
where \(\eta = 0.55\) for the FROG pulse and \(\eta = 0.46\) for the AFROG pulse.
\paragraph*{Time-evolution}
The time evolution operator \(U_{ij}(t_g)\) in the cost function is obtained by numerically simulating the system dynamics for each parameter pair \((\delta_i, \gamma_j)\).
The total gate time \(t_g\) is discretized into \(N\) time steps such that \(t_g =  N \Delta t\).
The control pulse is assumed to be piecewise constant over each interval, allowing the Hamiltonian to be treated as time-independent within each step.
At each time step \(n\), the propagator is given by \(P_{ij,n} = \exp(-i H_{ij,n} \Delta t)\), where \(H_{ij,n}\) is the Hamiltonian at that interval under parameter deviations ($\delta_i$, $\gamma_j$) defined in Eq.~\eqref{eq:Wright-Hamiltonian} of the main text.
The full time evolution operator is then constructed by sequentially multiplying these propagators:
\[
U_{ij}(t_g) = P_{ij,N} P_{ij,N-1} \cdots P_{ij,1}.
\]
\paragraph*{GRAPE for analytically-shaped pulses.}
To perform the GRAPE optimization, we require the gradients of the cost function with respect to the control functions, specifically \(\partial J / \partial \Omega_{x,n}\) and \(\partial J / \partial \Omega_{y,n}\) at each time step \(t_n\). These gradients guide the update of the control fields to minimize the cost function \(J\). For details on the gradient calculation methods used in GRAPE, we refer the reader to Refs.~\cite{Khaneja2005,Machnes2011,DeFouquieres2011}. In our case, we have:
\[\begin{aligned}
    &\frac{\partial J}{\partial \Omega_{x,n}}=\frac{1}{N_{\delta}N_{\gamma}}\sum_{i,j}\left(1+\tfrac{\gamma_j}{\Omega_0}\right)\mathbb{R}\left[\Tr\left(V_{ij,n}^{\dagger}X_{ij,n}U_{ij,n}\right)\right],\\
    &\frac{\partial J}{\partial \Omega_{y,n}}=\frac{1}{N_{\delta}N_{\gamma}}\sum_{i,j}\left(1+\tfrac{\gamma_j}{\Omega_0}\right)\mathbb{R}\left[\Tr\left(V_{ij,n}^{\dagger}Y_{ij,n}U_{ij,n}\right)\right],
\end{aligned}\]
where:
\begin{itemize}
    \item $U_{ij,n} = P_{ij,n} P_{ij,n-1}\cdots P_{ij,1}$,
    \item $V_{ij,n}^{\dagger}=V_{ij,N}^{\dagger}P_{ij,N}\cdots P_{ij,n}$ with:
    
    $\begin{aligned}
        V_{ij,N}=&-\tfrac{1}{2}U_{\mathrm{T}}\big(\ket{0}\bra{0}+\ket{1}\bra{1}\big)\times\\
        &\left(\bra{0} U_\text{T}^\dagger U_{ij}(t_g) \ket{0} + \bra{1} U_\text{T}^\dagger U_{ij}(t_g) \ket{1}\right),
    \end{aligned}
    $
    \item $X_{ij,n}=-i\frac{\Delta t}{2}\left(\hat{\sigma}^x_{10}+\hat{\sigma}^x_{21}\right)-\frac{\Delta t^2}{4}[H_{ij,n},\hat{\sigma}^x_{10}+\hat{\sigma}^x_{21}]+O(\Delta t^3)$,
    \item $Y_{ij,n}=-i\frac{\Delta t}{2}\left(\hat{\sigma}^y_{10}+\hat{\sigma}^y_{21}\right)-\frac{\Delta t^2}{4}[H_{ij,n},\hat{\sigma}^y_{10}+\hat{\sigma}^y_{21}]+O(\Delta t^3)$,
\end{itemize}

In our approach, the control pulse is not optimized directly at each time step. Instead, it is decomposed into a Fourier series, as defined in Eq.~\eqref{eq:Wright-FourierPulse}, and the optimization is performed over the corresponding Fourier coefficients \(a_n\) and \(b_n\), which determine the pulse shape. Consequently, we require the gradients \(\partial J / \partial a_n\) and \(\partial J / \partial b_n\), which can be computed using the chain rule~\cite{Skinner2010}:
\[
\begin{aligned}
    \frac{\partial J}{\partial a_n} &= \frac{\partial J}{\partial \Omega_{x,n}} \frac{\partial \Omega_{x,n}}{\partial a_n} + \frac{\partial J}{\partial \Omega_{y,n}} \frac{\partial \Omega_{y,n}}{\partial a_n}, \\
    \frac{\partial J}{\partial b_n} &= \frac{\partial J}{\partial \Omega_{x,n}} \frac{\partial \Omega_{x,n}}{\partial b_n} + \frac{\partial J}{\partial \Omega_{y,n}} \frac{\partial \Omega_{y,n}}{\partial b_n}.
\end{aligned}
\]
The gradients of the constraints in Eq.~\eqref{eq:PulseConstr}, which may be provided to the optimization routine to speed up the computation, are also obtained via this chain rule.
 
Once the cost function and its gradients are known, any standard optimization routine can be used to search for the set of parameters \(a_n\) and \(b_n\) that minimize the cost function in Eq.~\eqref{eq:GrapeCostFun2}. Examples of such routines are \texttt{fmincon} in \textsc{Matlab} or \texttt{scipy.optimize.minimize} in Python.

\paragraph*{Results.} In this work, we design pulses that implement robust \(X_{\pi/2}\) gates by setting the target unitary to \[U_{\text{T}} = \begin{pmatrix}
    \frac{1}{\sqrt{2}} & -\frac{i}{\sqrt{2}} & 0 \\
    -\frac{i}{\sqrt{2}} & \frac{1}{\sqrt{2}} & 0 \\
    0 & 0 & 1
\end{pmatrix}.\] To account for leakage effects in the simulations, the system includes the lowest three energy levels. In all cases, we use a time step of \(\Delta t = 0.5\,\text{ns}\) which corresponds to the arbitray waveform generator (AWG) sampling time. 

For the FROG pulse, we choose \(N = 224\) time steps, resulting in a total pulse duration of \(t_g = 112\,\text{ns}\). The pulse is optimized over a range of detunings \(\delta/(2\pi) \in [-0.5, 0.5]\)\,MHz, discretized into \(N_\delta = 21\) values. The amplitude scaling factor is fixed at \(\gamma = 0\), with \(N_\gamma = 1\). 

The AFROG pulse is designed with \(N = 256\) time steps, corresponding to a total duration of \(t_g = 128\,\text{ns}\). It is optimized over a range of detunings \(\delta / (2\pi) \in [-0.25, 0.25]\)~MHz and amplitude error \(\gamma/(2\pi) \in [-1.2, 1.2]\)~MHz, with \(N_\delta = N_\gamma = 21\) discrete values. The resulting pulse coefficients are given in Tab.~\ref{tab:Fourier_coeff}, where the pulse envelopes are shown in Fig.\,\ref{fig:landscapes}~(a) and (b) in the main text.
\begin{table}[H]
    \centering
    \begin{tabular}{|ll|ll|}
    \hline
         \multicolumn{2}{|c|}{FROG}& \multicolumn{2}{c|}{AFROG} \\
         \hline
        $a_1=-0.6137$ & $b_1=-0.0106$ & $a_1= 0.3492$ & $b_1= 0.0859$ \\
        $a_2=-0.0247$ & $b_2= 0.0334$ & $a_2=-0.2470$ & $b_2=-0.4581$ \\
        $a_3= 0.0742$ & $b_3= 0.0579$ & $a_3=-0.2474$ & $b_3=-0.0206$ \\
        $a_4= 0.0507$ & $b_4= 0.0140$ & $a_4=-0.0773$ & $b_4= 0.0213$ \\
        $a_5= 0.0149$ & $b_5=-0.0416$ & $a_5= 0.0352$ & $b_5= 0.0614$ \\
         \hline
    \end{tabular}
    \caption{Fourier coefficients of the optimal pulses. The pulses are given by Eq.~\eqref{eq:Wright-FourierPulse}}
    \label{tab:Fourier_coeff}
\end{table}

%% file: Appendices/appC_calibration.tex
\section{\texorpdfstring{\\* \vspace{2mm}}~Gate calibration}\label{app:cal}
\begin{figure*}[t]
\includegraphics[width=\textwidth]{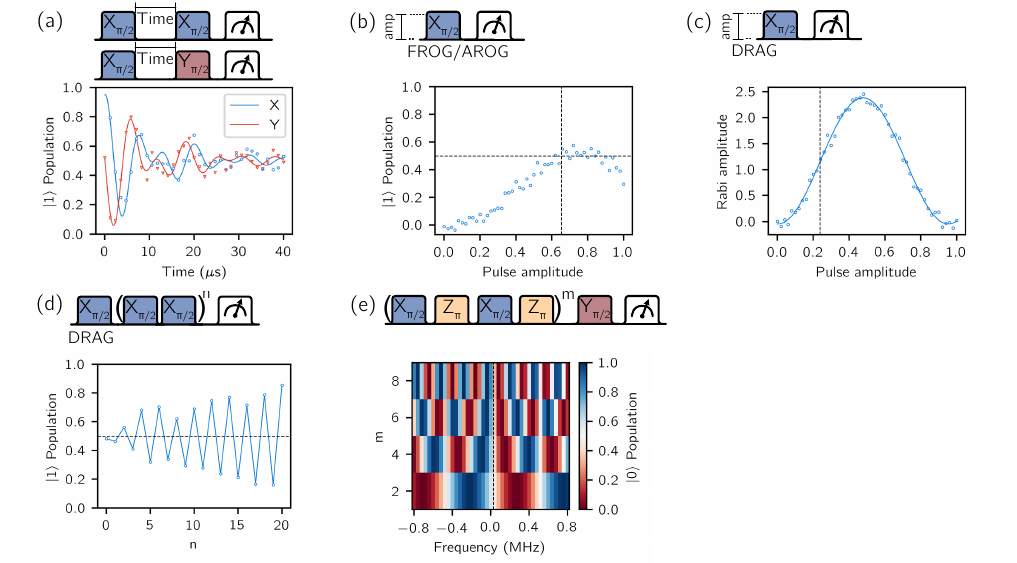}
\caption{\label{fig:calibration} 
Tune-up routine for single qubit gates.
The qubit frequency is calibrated using a Ramsey experiment (a).
Gates are then calibrated with a simple amplitude sweep, with FROG and AROG in (b) and DRAG in (c).
Subsequently, the DRAG gate is fine-calibrated using error amplification sequences for the pulse amplitude (d), and frequency and DRAG parameter (e).}
\end{figure*}
Each pulse parameter is calibrated sequentially to fully tune-up the gates.
We first characterize the qubit frequency $\omega_{01}$ using a Ramsey pulse sequence (see Fig\,\ref{fig:calibration}\,(a)).
Due to charge partiy switching, we see two frequencies leading to a beating pattern.
We take the qubit frequency $\omega_{01}$ to be the average of these two frequencies.

We proceed to sweep the pulse amplitude for all gates until the desired population is reached.
An $X_{\pi/2}$ pulse is realized when the population reaches 0.5, as shown in Fig.\,\ref{fig:calibration}\,(b) for FROG and AROG, and (c) for DRAG.
Note that the qubit rotation angle does not change linearly with the pulse amplitude for the robust gates, and the desired population lies on a small plateau.

For the DRAG gate, we perform additional tuning with fine amplitude targeting and simultaneous precise calibrations of the drive frequency and the DRAG correction parameter.
Error amplification sequences are performed to iteratively fine-tune each pulse parameter.
For amplitude, the sequence $(X_{\pi/2})^{2n + 1}$ is measured, as shown in Fig.\,\ref{fig:calibration}\,(d).
The final population is fitted to 
\begin{equation}\label{eq:Wright-ErrorAmp}
    P(\ket0) = a + \frac12 (-1)^n \cos\left(\frac{\pi}{2} + n\pi\frac{\gamma}{\Omega_0}\right)
\end{equation}
as a function of the number of repetitions $n$, where $a$ is left as a fitting parameter which goes to $1/2$ for a perfect $X_{\pi/2}$ gate and $\gamma$ is the amplitude error as defined in Eq.~\eqref{eq:Wright-Hamiltonian}~\cite{Sheldon2016}.
The qubit undergoes a drive-induced frequency shift, which we also calibrate for the DRAG gate.
The frequency is fine-tuned using the sequence $(X_{\pi/2}Z_\pi X_{\pi/2}Z_\pi)^{m}Y_{\pi/2}$ for various $m$~\cite{Sheldon2016}, see Fig.\,\ref{fig:calibration}\,(f). 
The DRAG parameter $\beta$ in
\begin{equation}
    \Omega(t) = \Omega_x(t) + i\beta \dot\Omega_x(t)
\end{equation}
is tuned using the same sequence, after the frequency offset calibration.